\documentclass[pra,superscriptaddress,amsmath,amssymb,12pt]{revtex4}
\usepackage{amsthm}
\usepackage{graphicx}

\newcommand{\Prf}{\noindent\textbf{Proof.\ }}

\newcommand{\bra}[1]{\langle #1 |}
\newcommand{\ket}[1]{| #1 \rangle}
\newcommand{\braket}[2]{\langle #1 | #2 \rangle}
\newcommand{\ketbra}[2]{\ket{#1}\bra{#2}}
\newcommand{\proj}[1]{\ket{#1}\bra{#1}}
\newcommand{\tr}{\text{Tr}}
\newcommand{\one}{{\bf 1}}

\newcommand{\Hil}{\mathcal H}
\newcommand{\channel}{\mathcal E}
\newcommand{\channelb}{\mathcal E^\dagger}

\newcommand{\cref}[1]{(\ref{#1})}

\newcommand{\zeroset}[1]{{\mathbb K}}

\newtheorem{thm}{Theorem}
\newtheorem{prop}[thm]{Proposition}
\newtheorem{lem}[thm]{Lemma}
\newtheorem{dfn}[thm]{Definition}

\newtheorem{cor}[thm]{Corollary}

\begin{document}

\title{Quantum Error Correction of Observables}
\author{C\'edric B\'eny}
\affiliation{Department of Applied Mathematics, University of
Waterloo, ON, Canada, N2L 3G1}
\author{Achim Kempf}
\affiliation{Department of Applied Mathematics, University of
Waterloo, ON, Canada, N2L 3G1}
\author{David~W.~Kribs}
\affiliation{Department of Mathematics and Statistics, University
of Guelph, Guelph, ON, Canada, N1G 2W1} \affiliation{Institute for
Quantum Computing, University of Waterloo, ON Canada, N2L 3G1}
\date{\today}
\begin{abstract}
A formalism for quantum error correction based on operator
algebras was introduced in \cite{beny07} via consideration of the
Heisenberg picture for quantum dynamics. The resulting theory
allows for the correction of hybrid quantum-classical information
and does not require an encoded state to be entirely in one of the
corresponding subspaces or subsystems. Here, we provide detailed
proofs for the results of \cite{beny07}, derive a number of new
results, and we elucidate key points with expanded discussions. We
also present several examples and indicate how the theory can be
extended to operator spaces and general positive operator-valued
measures.
\end{abstract}
\maketitle

\section{Introduction}\label{S:intro}

A new framework for quantum error correction was derived in
\cite{beny07} through a Heisenberg picture reformulation of the
Schr\"odinger approach to error correction, and an expansion of
the notion of a quantum code to allow for codes determined by
algebras generated by observables. As the approach generalizes
standard quantum error correction (QEC)
\cite{bennett96,knill97,shor95, steane96, gottesman96} and
operator quantum error correction (OQEC) \cite{kribs05, kribs06},
we called the resulting theory ``operator algebra quantum error
correction'' (OAQEC). An important feature of OAQEC is that it
provides a formalism for the correction of hybrid
quantum-classical information \cite{Kuper03}.

In this paper we provide proofs for the results stated in
\cite{beny07}, and we establish a number of new results. In
addition, we expose some of the finer points of the theory with
discussions and several examples. We also outline how the theory
can be extended to the case of operator spaces generated by
observables and general positive operator-valued measures (POVMs).

We continue below by establishing notation and describing
requisite preliminary notions. In the next section we present a
detailed analysis of passive quantum error correction within the
OAQEC framework. The subsequent section does the same for active
quantum error correction. This is followed by an expanded
discussion of the application to information flow from
\cite{beny07}, and we conclude with a section on the operator
space and POVM extension.

\subsection{Preliminaries}\label{sS:prelims}

Given a (finite-dimensional) Hilbert space $\Hil$, we let
$\mathcal L(\Hil)$ be the set of operators on $\Hil$ and let
$\mathcal L_1(\Hil)$ be the set of density operators on $\Hil$. We
shall write $\rho$, $\sigma$, $\tau$ for density operators and
$X$, $Y$, etc, for general operators. The identity operator will
be written as $\one$.

Noise models in quantum computing are described  (in the
Schr\"odinger picture) by completely positive trace-preserving
(CPTP) maps $\mathcal E: \mathcal L(\Hil)\rightarrow \mathcal
L(\Hil)$ \cite{NC00}. We shall use the term {\it quantum channel}
to describe such maps. Every map $\mathcal E$ has an operator-sum
representation $\mathcal E(\rho) = \sum_a E_a \rho E_a^\dagger$,
where the operators $E_a$ are called the {\it operation elements}
or {\it noise operators} for $\mathcal E$. The Hilbert-Schmidt
dual map $\mathcal E^\dagger$ describes the corresponding
evolution of observables in the Heisenberg picture. A set of
operation elements for $\mathcal E^\dagger$ is given by
$\{E_a^\dagger\}$. Trace preservation of $\mathcal E$ is
equivalent to the requirement that $\mathcal E^\dagger$ is {\it
unital}; that is, $\mathcal E^\dagger (\one) = \one$.

A quantum system $A$ (or $B$) is a subsystem of $\Hil$ if $\Hil$
decomposes as $\Hil = (A\otimes B) \oplus \mathcal K$. Subspaces
of $\Hil$ can clearly be identified as subsystems $A$ with
one-dimensional ancilla ($\dim B =1$). An algebra $\mathcal A$ of
operators on $\Hil$ that is closed under Hermitian conjugation is
called a (finite-dimensional) C$^*$-algebra, what we will simply
refer to as an ``algebra''. Algebras of observables play a key
role in quantum mechanics \cite{vN55} and recently it has been
shown that they can be used to encode hybrid quantum-classical
information \cite{Kuper03}. Below we shall discuss further the
physical motivation for considering algebras in the present
setting. Mathematically, finite-dimensional C$^*$-algebras have a
tight structure theory that derives from their associated
representation theory \cite{Dav96}. In particular, there is a
decomposition of $\Hil$ into subsystems $\Hil = \oplus_k
(A_k\otimes B_k) \bigoplus \mathcal K$ such that with respect to
this decomposition the algebra is given by
\begin{eqnarray}\label{algeqn}
\mathcal A = \Big[ \bigoplus_k \big( \mathcal L(A_k) \otimes
\one_{B_k} \big)\Big] \,\, \bigoplus \,\, 0_{\mathcal K}.
\end{eqnarray}
The algebras $\mathcal L(A_k) \otimes \one_{B_k}$ are referred to
as the ``simple'' sectors of $\mathcal A$. We shall write
$\mathcal M_n$ for the set of $n\times n$ complex matrices, and
identify $\mathcal M_n$ with the matrix representations for
elements of $\mathcal L(A)$ when $\dim A=n$ and an orthonormal
basis for $A$ is fixed.

\section{Passive Error Correction Of Algebras}\label{S:passive}

As the terminology suggests, the existence of a passive code for a
given noise model implies that no active operation is required
(beyond decoding) to recover quantum information encoded therein.
Mathematically,  it is quite rare for a generic channel to have
passive codes. However,  many of the naturally arising physical
noise models include symmetries that do allow for such codes
\cite{zanardi97, palma96, duan97, lidar98,knill00, zanardi01a,
kempe01, choi06, knill06, holbrook04, holbrook05, junge05}.

The following is the standard definition of a noiseless subsystem
(and decoherence-free subspace when $\dim B =1$) in the
Schr\"odinger picture. Suppose we have a decomposition of the
Hilbert space $\Hil$ as $\Hil = (A \otimes B) \oplus \mathcal K$.
As a notational convenience, we shall write $\rho \otimes \tau$
for the operator on $\Hil$ defined by $(\rho\otimes\tau)\oplus
0_{\mathcal K}$.

\begin{dfn}\label{nsdefn}
We say that $A$ is a noiseless (or decoherence-free) subsystem for
$\channel$ if for all $\rho \in \mathcal L_1(A)$ and $\sigma \in
\mathcal L_1(B)$ there exists $\tau \in \mathcal L_1(B)$ such that
\begin{eqnarray}\label{nseqn}
\channel(\rho \otimes \sigma) = \rho \otimes \tau.
\end{eqnarray}
\end{dfn}

The ``decoherence-free'' terminology is usually reserved for
subspaces (when $\dim B =1$).

\subsection{Decoherence-Free And Noiseless Subspaces And Subsystems In The \\ Heisenberg Picture}\label{sS:dfs/ns}

The following theorem gives an equivalent formulation of this
definition in the Heisenberg picture, that is in terms of the
evolution of observables, given by the dual channel $\channelb$.
We introduce the projector $P$ of $\Hil$ onto the subspace $A
\otimes B$.

\begin{thm}\label{nsthm}
$A$ is a noiseless subsystem for $\mathcal E$ if and only if
\begin{equation}
\label{ns} P\,\channelb(X\otimes \one)\,P = X \otimes \one
\end{equation}
for all operators $X\in\mathcal L(A)$
\end{thm}

\Prf If $A$ is a noiseless subsystem for $\mathcal E$, then for
all $X \in \mathcal L(A)$ we have
\begin{eqnarray*}
\tr(P{\mathcal E}^\dagger(X\otimes \one) P(\rho\otimes\sigma)) &=& \tr(\channelb(X\otimes \one) (\rho \otimes \sigma)) = \tr((X\otimes \one) \channel(\rho \otimes \sigma))\\
& =& \tr(X\rho \otimes \tau) = \tr(X \rho) \tr(\tau) \\ &=& \tr(X
\rho) = \tr((X \otimes \one) (\rho \otimes \sigma)).
\end{eqnarray*}
This is true for all $\rho\in \mathcal L_1(A)$ and all $\sigma \in
\mathcal L_1(B)$. By linearity it follows that
$P\channelb(X\otimes \one)P = X \otimes \one$ for all $X \in
\mathcal L(A)$.

Reciprocally, if we assume Eq.~(\ref{ns}) to be true for all $X
\in \mathcal L(A)$, then for all $\gamma\in \mathcal L_1(A \otimes
B)$ we have
\begin{eqnarray*}
\tr(X\, \tr_B(P\channel(\gamma)P))
&=& \tr((X\otimes \one) P\channel(\gamma)P) = \tr(P \channelb(X\otimes \one) P \gamma) \\
&=& \tr((X\otimes \one) \gamma ) = \tr(X \,\tr_B(\gamma)),
\end{eqnarray*}
where we have freely used the facts $P(X\otimes\one)P = X\otimes
\one$ and $P\gamma P = \gamma$. Since the above equation is true
for all $X$, we have $\tr_B(P\channel(\gamma)P) = \tr_B(\gamma)$
for all $\gamma\in\mathcal L_1(A\otimes B)$, which was shown in
\cite{kribs06} to be equivalent to the definition of $A$ being a
noiseless subsystem for $\mathcal E$. \qed

\strut

Note that Eq.~(\ref{ns}) can be satisfied even if part of an
observable $X\otimes \one$ spills outside of the subspace $P\Hil$
under the action of $\mathcal E^\dagger$. The projectors $P$ in
Eq.~(\ref{ns}) show that the noiseless subsystem condition in the
Heisenberg picture is only concerned with the ``matrix corner'' of
$\mathcal E^\dagger(X\otimes \one)$ partitioned by $P$.


\strut

In some cases the equivalence of Eq.~(\ref{nseqn}) and
Eq.~(\ref{ns}) can be seen from a different perspective. For a
bistochastic or unital channel (those for which $\mathcal
E(\one)=\one$), the dual $\mathcal E^\dagger$ is also a channel.
Then structural results for unital channels from \cite{kribs03}
can be used to give an alternate realization of this equivalence,
and passive codes may be computed directly from the commutant of
the operation elements for $\mathcal E$. The simplest case would
be for ``self-dual'' channels, those for which $\mathcal E^\dagger
= \mathcal E$. Clearly, any $\mathcal E$ with Hermitian operation
elements is self-dual. In particular, self-dual channels include
all Pauli noise models, which are channels with operation elements
belonging to the Pauli group, the group generated by tensor
products of unitary Pauli operators $X$, $Y,$ $Z$.

As a further illustrative (non-unital) example, consider the
single-qubit spontaneous emission channel \cite{NC00} given by
$\mathcal E(\sigma) = \tr(\sigma)\, \ket{0}\bra{0}$. Here $P =
\ket{0}\bra{0}$. This channel is implemented by operation elements
$E_0 = P$ and $E_1 = \ket{0}\bra{1}$. Hence the dual channel is
given by $\mathcal E^\dagger(\sigma) = P\sigma P + E_1^\dagger
\sigma E_1$. The subspace $A$ spanned by the ground state
$\ket{0}$ is a decoherence-free subspace for $\mathcal E$ (though
it cannot be used to encode quantum information since $\dim A
=1$). In this case, Eq.~(\ref{ns}) is equivalent to the statement
$P \mathcal E^\dagger (P) P = P$, which may be readily verified.

One can consider more general spontaneous emission channels with
non-trivial decoherence-free subspaces. For instance, consider a
qutrit noise model that describes spontaneous emission from the
second excited state to the ground state. The corresponding
channel is defined by $\mathcal E(\sigma) = P\sigma P + E \sigma
E^\dagger$, where $P = \ket{0}\bra{0} + \ket{1}\bra{1}$ and $E =
\ket{0}\bra{2}$. The subspace $P \mathbb{C}^3 = {\rm
span}\,\{\ket{0},\ket{1}\}$ is a single-qubit decoherence-free
subspace for $\mathcal E$ since $\mathcal E(\sigma)= \sigma$ for
all $\sigma = P\sigma P$. It is also easy to see in this case that
Eq.~(\ref{ns}) is satisfied since $P \mathcal E^\dagger(X)P = X$
for all $X\in \mathcal L(P\mathbb{C}^3)$. Notice also in this
example that $\mathcal E^\dagger$ can induce ``spillage'' from $P$
to $P^\perp$. Indeed, this can be seen immediately from the
operator-sum representation for $\mathcal E^\dagger$;
specifically, for all $X\in \mathcal L(P\mathbb{C}^3)$ we have
\begin{equation}
\mathcal E^\dagger(X) = PXP + E^\dagger X E = X +  \alpha\,
\ket{2} \bra{2},
\end{equation}
where $\alpha = \bra{0} X \ket{0}$.

\subsection{Conserved Algebras Of Observables}\label{sS:conserved}


A POVM determined by a set of operators $X = \{X_a\}$ evolves via
the unital CP-map $\channelb$ in
the Heisenberg picture. If for all $a$ we have $X_a = \channelb(X_a)$, then 
all the statistical information about  $X = \{X_a\}$ has been
conserved. Indeed, for any initial state $\rho$, we have $\tr(\rho
X_a) = \tr(\rho \mathcal E^\dagger(X_a)) =
\tr(\channel(\rho)X_a)$. Moreover, if we have control on the
initial states, an expected feature in quantum computing, we can
ask which elements are conserved if the state starts in a certain
subspace $P\Hil$; that is, which elements satisfy $P\channelb
(X_a)P = PX_aP$, or equivalently $\tr(X_a \,\mathcal E(P\rho P)) =
\tr(X_a \, P\rho P)$ for all $\rho \in \mathcal L_1(\Hil)$. This,
together with the Heisenberg characterization of Eq.~(\ref{ns}),
motivates the following definition.


\begin{dfn}\label{conserved}
We shall say that a set $\mathcal S$ of operators on $\Hil$ is
{\em conserved} by $\channel$ for states in $P\Hil$ if every
element of $\mathcal S$ is conserved; that is, if
\begin{equation}\label{conservedalg}
P\,\channelb(X)\,P = PXP \quad \forall X \in \mathcal S.
\end{equation}
\end{dfn}

The focus of the present work is  error correction for algebras
generated by observables. Let us  consider in more detail the
physical motivation for considering algebras. In the Heisenberg
picture a set of operators $\{X_a\}$ evolves according to the
unital CP-map $\channelb$ with elements $E_a^\dagger$ instead of
$E_a$. If for all values of the label $a$ we have $X_a =
\channelb(X_a)$ then all the statistical information about $X$ has
been conserved by $\mathcal E$ as noted above. In such a scenario
we say that $X_a$ is conserved by $\channel$. In particular, if
$X$ defines a standard projective measurement, $X = \sum_a p_a
X_a$ with $X_a^2 = X_a$ for all $a$, then the projectors $X_a$
linearly span the algebra they generate. Hence, in this case
$\channel$ conserves an entire commutative algebra. Therefore,
focussing on the conservation (and more generally correction
defined later) of sets of operators that have the structure of an
algebra,
apart from allowing a complete characterization, is also
sufficient for the study of all the correctable {\em projective}
observables.

Further observe that Eq.~(\ref{conservedalg}) applied to a set of
observables that generate an (arbitrary) algebra gives a
generalization of noiseless subsystems. Indeed, by
Theorem~\ref{nsthm} any subalgebra $\mathcal A$ of $\mathcal
L(P\Hil)$ for which all elements $X\in \mathcal A$ satisfy
Eq.~(\ref{conservedalg}) is a direct sum of simple algebras, each
of which encodes a noiseless subsystem (when $\dim A_k >1$ as in
Eq.~(\ref{algeqn})). It is also important to note that given an
algebra $\mathcal A$ conserved for states in $P\Hil$ (or more
generally correctable as we shall see) quantum information cannot
in general be encoded into the entire subspace $P\Hil$ for safe
recovery, but rather into subsystems of $P\Hil$ determined by the
splitting induced from the algebra structure of $\mathcal A$.
These points will be further expanded upon in the discussion of
Section~\ref{sS:schrodinger}.

We now establish concrete testable conditions for passive error
correction. Namely, these conditions  are stated strictly in terms
of the operation elements for a channel. This result was stated
without proof in \cite{beny07}. The special case of simple
algebras in the Schr\"odinger picture was obtained in
\cite{kribs05,kribs06}. Techniques of \cite{choi06} are used in
the analysis. We first present a simple lemma that will be used
below.

\begin{lem}
\label{lemm} Let $\mathcal F$ be a CP map with elements $F_a$. If
$A \ge 0$ is such that $\mathcal F (A) = 0$ then it follows that
$AF_a = 0$ for every element $F_a$.
\end{lem}

{\noindent}{\bf Proof.} If $\sum_a F_a^\dagger A F_a  = 0$ then
for any state $\ket{\psi}$, $ \sum_a \bra{\psi} F_a^\dagger A F_a
\ket{\psi} = 0. $ Since each operator $F_a^\dagger A F_a$ is
positive, this is a sum of nonnegative terms. Therefore each
individual term must equal zero; $ \bra{\psi} F_a^\dagger A F_a
\ket{\psi} = 0 $ for all $a$. This means that the vector $\sqrt{A}
F_a \ket{\psi}$ is of norm zero, and therefore is the zero vector.
This being true for all states $\ket{\psi}$, we must have that
$\sqrt{A} F_a = 0$, from which it follows that $A F_a = 0$. \qed

\begin{thm}
\label{cons} Let $\mathcal A$ be a subalgebra  of $\mathcal
L(P\Hil)$. The following statements are equivalent:
\begin{enumerate}
\item $\mathcal A$ is conserved by $\mathcal E$ for states in
$P\Hil$. \item $[E_a P, X] = 0$ for all $X\in \mathcal A$ and all
$a$.
\end{enumerate}
\end{thm}

{\noindent}{\bf Proof.} If $[E_aP,X]=[E_a P, PXP] = 0$ for all
$a$, then
\begin{eqnarray*}
P\channelb(X)P = \sum_a P E_a^\dagger X E_a P &=& \sum_a P
E_a^\dagger P X P E_a P \\ &=& \sum_a P E_a^\dagger E_a P X P =
P{\mathcal E}^\dagger(\one) P X P = P X P = X.
\end{eqnarray*}

Reciprocally, we assume that each $X \in \mathcal A$ satisfies
$P\channelb(X)P = P X P = X$.
Consider a projector $Q \in  \mathcal A $. We have $P\channelb(Q)P
= Q$ and hence $P\channelb(Q^\perp )P = P Q^\perp$. Therefore
\[
Q^\perp P \channelb(Q) P Q^\perp = Q^\perp\, Q\, Q^\perp = 0,
\]
and similarly $Q P \channelb(Q^\perp ) P Q = 0$. By
Lemma~\ref{lemm}, this implies respectively $Q E_a P Q^\perp = 0$
and $Q^\perp E_a P Q = 0$ for all $a$. Together these conditions
imply
\[
QE_a P = Q E_a P Q = E_a PQ,
\]
 and thus $[E_a P, Q] = 0$ for all $a$. Finally, note that since $ \mathcal A $ is an
algebra, then a generic element $Y \in  \mathcal A $ can be
written as a linear combination of projectors in $ \mathcal A $.
Therefore we also have $[E_a P, Y] = 0$ for all $Y \in  \mathcal
A$, and this completes the proof. \qed

\smallskip

This theorem allows us to identify the largest conserved
subalgebra of $\mathcal L(P\Hil)$ conserved on the subspace
$P\Hil$. Namely, a direct consequence of Theorem~\ref{cons} is
that the (necessarily $\dagger$-closed) commutant inside $\mathcal
L(P\Hil)$ of the operators $\{E_aP,PE_a^\dagger\}$ is the largest
such algebra.

\begin{cor} \label{largestalgebra}
The algebra $\mathcal A = \{X\in\mathcal L(P\Hil) : \forall
a\,\,[X,E_aP] = [X^\dagger,E_a P] = 0 \}$
is conserved on states in $P\Hil$ and contains all subalgebras of
$\mathcal L(P\Hil)$ conserved on states $P\Hil$.
\end{cor}

Note that $P$ itself may not belong to this algebra, unless it
satisfies $E_a P = P E_a P$.  This special case was the case
considered in \cite{choi06}. With other motivations in mind, the
special case $P=\one$ was also derived in \cite{lindblad99} where
it was shown that the full commutant of $\{E_a,E_a^\dagger\}$ is
the largest algebra inside the fixed point set of a unital CP map.
(This in turn may be regarded as a weaker form of the fixed point
theorem for unital channels \cite{kribs03}.)

\strut

Likely the most prevalent class of decoherence-free subspaces are
the stabilizer subspaces for abelian Pauli groups, which give the
starting point for the stabilizer formalism \cite{gottesman96}. On
the $n$-qubit Hilbert space, let $s<n$ and consider the group $S$
generated by the Pauli phase flip operators $Z_1,\ldots ,Z_s$,
where we have written $Z_1 = Z\otimes \one\otimes \ldots \otimes
\one$, etc. The joint eigenvalue-1 space for $S$ is a
$2^{n-s}$-dimensional decoherence-free subspace for $\{Z_j:1\leq j
\leq s\}$, called the ``stabilizer subspace'' for $S$. In the
stabilizer formalism, any of the $2^s$ (necessarily
$2^{n-s}$-dimensional) mutually orthogonal joint eigenspaces for
elements of $S$ could be used individually to build codes, but the
eigenvalue-1 space is used as a convenience. Each of these
eigenspaces supports a full matrix algebra $\mathcal M_k$, where
$k=2^{n-s}$. In the present setting these decoherence-free
subspaces may be considered together, as they are defined by the
structure of the commutant $S^\prime \cong \mathcal M_k^{(2^s)}$.
In particular, this entire algebra is conserved by any channel
defined with operation elements given by linear combinations of
elements taken from $S$.

Allowing other Pauli operators into the error group can result in
non-conserved scenarios. For instance, consider the case $n=3$ and
$s=2$. The algebra $S^\prime \cong \mathcal M_2^{(4)}$ is
conserved by channels determined by elements of $S$ as above.
However, it is {\it not} conserved by channels determined by
elements of $G = \{Z_1,Z_2,X_1X_2\}$, even though the individual
eigenvalue-1 stabilizer space is still a correctable code for $G$.
This follows from Theorem~\ref{cons} since $G$ properly contains
$S$, and hence $G^\prime$ is properly contained in $S^\prime$. (In
this case $P=\one$, and the largest conserved algebra is the full
commutant $G^\prime$ of the error group.) Interestingly, there are
still algebras conserved by channels determined by elements of
$G$, since its commutant has the structure $G^\prime = {\rm
Alg}\{G\}^\prime \cong (I_2\otimes\mathcal
M_2)\oplus(I_2\otimes\mathcal M_2)$.

\strut


Let us reconsider the special case in which the projector $P$
itself is one of the correctable observables. Most importantly,
this guarantees that after evolution all states are back into the
code $P\Hil$. Indeed, the probability that a state $\rho$
initially in the code is still in the code after evolution is then
given by
\begin{eqnarray}\label{projectorin}
p = \tr( \mathcal \channel(\rho) P) = \tr(\rho P
\channel^\dagger(P)P) = \tr(\rho P) = 1 .
\end{eqnarray}
We note that this also guarantees a repetition of the noise map
will  be conserved.

A refinement of the proof of Theorem~\ref{cons} yields the
following result when the projector $P$ is conserved and belongs
to the algebra of observables in question.

\begin{thm}
\label{cons2} Let $\mathcal A$ be an algebra containing the
projector $P$. The following statements are equivalent:
\begin{enumerate}
\item $\mathcal A$ is conserved by $\mathcal E$ for states in
$P\Hil$. \item $[E_a P, X] = 0$ for all $X\in P {\mathcal A} P$
and all $a$.
\end{enumerate}
\end{thm}

{\noindent}{\bf Proof.}  If $[E_aP,X]=[E_a P, PXP] = 0$ for all
$a$, then $P\channelb(X)P = PXP$ follows as above. On the other
hand, if each $X \in \mathcal A$ satisfies $P\channelb(X)P = P X
P$, then note that for $X = P$ this yields $P\channelb(P)P = P$
which implies $P\channelb(P^\perp)P = 0$. Therefore, by Lemma
\ref{lemm}, $P^\perp E_a P = 0$, that is, $E_a P = P E_a P$ for
all $a$. Hence for all $X \in \mathcal A$ we also have $
P\channelb(PXP)P = PXP. $ But the set $P\mathcal A P$ is itself an
algebra, since $P$ belongs to $\mathcal A$, and hence is spanned
by its projectors. The rest of the proof proceeds as above.
 \qed

\section{Active Error Correction Of Algebras}\label{S:correction}

More generally, active intervention into a quantum system may  be
required for error correction. In particular, we should be able to
protect a set of operators $\{X_a\}$ (which could define a POVM
for instance) by acting with  a channel $\mathcal R$ such that
each $X_a$ is mapped by $\mathcal R^\dagger$ to one of the
operators $Y_a = \channel^\dagger (X_a)$. That is, $\mathcal
R^\dagger(X_a) = Y_a$, and thus $(\mathcal R \circ
\channel)^\dagger(X_a) = X_a$. This, together with the previous
discussions, motivates the following definition.

\begin{dfn}\label{correctable}
We say that a set $\mathcal S$ of operators on $\Hil$ is {\em
correctable} for $\mathcal E$ on states in the subspace $P\Hil$ if
there exists a channel $\mathcal R$ such that $\mathcal S$ is
conserved by $\mathcal R \circ \channel$ on states in $P\Hil$; in
other words,
\begin{equation}\label{Heiseneqn}
P(\mathcal R \circ \channel)^\dagger(X)P = PXP \quad \forall
X\in\mathcal S.
\end{equation}
\end{dfn}

This equation is the same as the one we used to defined passive
error correction, except that we now allow for an arbitrary
``correction operation'' $\mathcal R$ after the channel has acted.
In particular, as in other settings, the passive case may be
regarded as the special case of active error correction for which
the correction operation is trivial, $\mathcal R = {\rm id}$.

This notion of correctability is more general than the one
addressed by the framework of OQEC, and extends the one introduced
in \cite{beny07} from algebras to arbitrary sets of observables.
Here we shall continue to focus on algebras, and in
Section~\ref{S:operatorspaces} we consider an extension of the
notion to general operator spaces. Whereas OQEC focusses on simple
algebras, $\mathcal L(A)\otimes\one^B$, here correctability is
defined for any set, and in particular for any finite-dimensional
algebra. See Section~\ref{sS:schrodinger} for an expanded
discussion on the form of OAQEC codes.

\subsection{Testable Conditions For OAQEC Codes}\label{sS:testable}


A set of operation elements for a given channel are the
fundamental building blocks for the associated physical noise
model. Thus, a characterization of a correctable OAQEC code
strictly in terms of the operation elements for a given channel is
of immediate interest. The following result was stated without
proof in \cite{beny07}. It generalizes the central result for both
QEC \cite{knill97} and OQEC \cite{kribs05, kribs06, nielsen05}.

\begin{thm}
\label{found} Let $\mathcal A$ be a subalgebra of $\mathcal
L(P\Hil)$. The following statements are equivalent:
\begin{enumerate}
\item $\mathcal A$ is correctable for $\mathcal E$ on states in
$P\Hil$.  \item $[P E_a^\dagger E_b^{} P, X] = 0$ for all $X \in
\mathcal A$ and all $a,b$.
\end{enumerate}
\end{thm}

{\noindent}{\bf Proof.} We write  $R_a$ for the elements of
$\mathcal R$. According to Theorem~\ref{cons}, the conservation of
$\mathcal A$ by $\mathcal R \circ \mathcal E$ implies $R_a E_b X =
X R_a E_b P$ for all $X \in \mathcal A$ and all $a,b$. But we also
have $R_a E_b X^\dagger = X^\dagger R_a E_b P$, so that $X
E_b^\dagger R_a^\dagger = P E_b^\dagger R_a^\dagger X$. Therefore
$P E_c^\dagger E_b X = \sum_a P E_c^\dagger R_a^\dagger R_a E_b X
= \sum_a P E_c^\dagger R_a^\dagger X R_a E_b P = X E_c^\dagger E_b
P.$

We will prove the sufficiency of this condition by explicitly
constructing a correction channel. For $k\geq 1$, let $P_k$ be the
projector onto the $k$th simple sector of the algebra $\mathcal
A$. Also let $P_0 = P - \one_{\mathcal A}$ where $\one_{\mathcal
A}$ is the unit element of the algebra $\mathcal A$. We have
$P_{k} E_b^\dagger E_c P_{k} = \one \otimes A_{bc}$ for some
operators $A_{bc}$ and all $k \geq 0$. Hence the theory of
operator error correction guarantees that each subspace can be
individually corrected. Here however we have the additional
property $P_{k} E_b^\dagger E_c P_{k'} = 0$ whenever $k \neq k'$,
which allows the correction of the state even if it is in a
superposition between several of the subspaces $P_k$.
Explicitly, we have $(\one \otimes \bra{l})P_k ( E_b^\dagger E_c)
P_{k'} (\one \otimes \ket{l'}) = \delta_{kk'}\lambda_{bc}^{kll'}
\one$ for some $\lambda_{bc}^{kll'} \in \mathbb C$, where we
denote $\one=\one_{m_k}$. According to the standard theory of
error correction this condition guarantees the existence of
channels $\mathcal R_k$ correcting the error operators $F_{ckl} =
E_c P_k (\one \otimes \ket{l})$ for all $l$ and all $c$, or any
linear combination of them. In particular, we will consider linear
combinations of the form
$\tilde F_{ck} = \sum_n \braket{n}{\psi} F_{ckn} =  E_c P_k (\one
\otimes \ket{\psi})$
for any normalized vector $\ket{\psi}$. Furthermore, from the
standard theory we know the elements of the correction channels
$\mathcal R_k$ can be assumed to have the form
$R^{(k)}_{cl} = \sum_{bj} \alpha^{(k)}_{clbj} (\one \otimes
\bra{j}) P_k E_b^\dagger$ for some complex numbers
$\alpha^{(k)}_{clbj}$.

We now show that the trace-decreasing channel $\mathcal R$ with
elements $ R_{kcl} = P_k (\one \otimes \ket{l}) R^{(k)}_{cl} $
corrects the algebra $\mathcal A$ on states $P\Hil$ for the
channel $\channel$.
First note that 
$R^{(k)}_{cl} E_a P = \sum_{bj}  \alpha^{(k)}_{clbj} (\one \otimes
\bra{j}) P_k E_b^\dagger E_a P_k = R^{(k)}_{cl} E_a P_k$. Hence,
for a general operator $X = \sum_k A_k \otimes \one$ in the
algebra we have $ P(\channelb \circ \mathcal R^\dagger)(X)P
= \sum_{akcl} P_k E_c^\dagger R^{(k)\dagger}_{al} A_k R^{(k)}_{al}
E_c P_k $. Considering each term $k$ separately, for any state
$\ket{\psi}$ we have
\[
\begin{split}
 \sum_{acl} (\one \otimes \bra{\psi}) &P_k E_c^\dagger R^{(k)\dagger}_{al} A_k R^{(k)}_{al} E_c P_k (\one \otimes \ket{\psi}) \\
&= A_k = (\one \otimes \bra{\psi})  (A_k \otimes \one) (\one
\otimes \ket{\psi}),
\end{split}
\]
where we have used the dual of the fact that $\mathcal R_k$
corrects the error operators $\tilde F_{ck}$. Therefore
$\sum_{acl} P_k E_c^\dagger R^{(k)\dagger}_{al} A_k R^{(k)}_{al}
E_c P_k = (A_k \otimes \one) $ and summing those terms over $k$
yields $P(\channelb \circ \mathcal R^\dagger)(X)P  = \sum_k (A_k
\otimes \one) = X$. \qed
\smallskip

As an immediate consequence of Theorem~\ref{found} we have the
following.

\begin{cor} \label{largestcorrectablealgebra}
The algebra $\mathcal A = \{X\in\mathcal L(P\Hil) : \forall
a,b\,\,[X,P E_a^\dagger E_bP] = 0 \}$
is correctable on states in $P\Hil$ and contains all subalgebras
of $\mathcal L(P\Hil)$ correctable on states in $P\Hil$.
\end{cor}

To further explain the structure  of the correction channel in
Theorem~\ref{found} let us show how it is constructed from OQEC
correction channels. This will also give an alternative proof of
the sufficiency of the correctability condition.

For simplicity, we will in fact build a channel which corrects the
larger algebra $\mathcal B := \mathcal A \oplus \mathbb C
(\one_\mathcal A - P)$. Remember that $P_k$ is the projector onto
the $k$th simple sector of the algebra $\mathcal A$, assuming a
decomposition as in Eq.~(\ref{algeqn}). Also we include $P_0 =
\one_\mathcal A - P$ which projects onto the additional sector in
$\mathcal B$.
Our correctability condition guarantees that each of those sectors
is an OQEC code. Let $\mathcal R_k$ be a OQEC correction channel
for the $k$th simple sector. We use the ``raw'' subunital version
of the subsystem correction channels whose elements are all linear
combinations of the operators $P_k E_a^\dagger $. They have the
property that $Q_k := \mathcal R_k^\dagger(\one) = \mathcal
R_k^\dagger(P_k)$ is a projector. Since the elements of the
channel $\mathcal R_k$ are linear combinations of the operators
$P_k E_a^\dagger$, we have $Q_k Q_l = \mathcal
R_k^\dagger(P_k)\mathcal R_l^\dagger(P_l) = 0$ if $k \neq l$,
because all the terms contain a factor of the form $P_k
E_a^\dagger E_b P_l = 0$. This means the the projectors $Q_k$ are
mutually orthogonal and sum to a projector $Q := \sum_k Q_k$. The
channels $\mathcal R_k$ also have the property that $\mathcal
R_k(\rho) = \mathcal R_k(Q_k \rho Q_k)$ for any state $\rho$. From
these ``local'' channels we can construct a trace-preserving
correction channel for the full algebra:
\begin{equation}
\mathcal R^\dagger(X) := \sum_k \mathcal R_k^\dagger(X) + \frac{\tr(PX)}{\tr(P)} Q^\perp 
\end{equation}
This CP map is a channel because $\mathcal R^\dagger(\one) =
\sum_k Q_k + Q^\perp = \one$. We have to check that it corrects
the algebra $\mathcal B$. First, concerning the effect of
$\channelb$ on the last term of the correction channel, note that
\[
\begin{split}
P \channelb( Q ) P &= \sum_k P \channelb( \mathcal R_k^\dagger(P_k)) P\\
 &= \sum_k P_k \channelb( \mathcal R_k^\dagger(P_k)) P _k = \sum_k P_k = P. \\
\end{split}
\]
Hence $P \channelb( Q^\perp ) P = 0$. It follows that for any $X
\in \mathcal B $ we have, keeping in mind that the channel
elements of $\mathcal R_k$ are linear combinations of the
operators $P_k E_a^\dagger$,
\[
\begin{split}
P \channelb( \mathcal R^\dagger(X)) P&= \sum_{kl} P_l \channelb( \mathcal R_k^\dagger(X)) P_l \\
 &= \sum_k P_k \channelb( \mathcal R_k^\dagger(X)) P_k \\
&= \sum_k P_k X P_k = X,
\end{split}
\]
which is the desired property for the correction channel $\mathcal
R$.
\smallskip

As in the passive case, we can consider the situation in which the
projector is correctable and belongs to the algebra. In this case,
the observables and states do not spill out from $P\Hil$ under the
action of $\mathcal R\circ\mathcal E$, and so the channel followed
by the correction operation is repeatable. The previous proof can
be readily refined for this purpose.

\begin{thm}
\label{found2} Let $\mathcal A$ be an algebra containing the
projector $P$. The following statements are equivalent:
\begin{enumerate}
\item $\mathcal A$ is correctable for $\mathcal E$ on states in
$P\Hil$. \item $[P E_a^\dagger E_b^{} P, X] = 0$ for all $X \in P
\mathcal A P$ and all $a,b$.
\end{enumerate}
\end{thm}

{\noindent}{\bf Proof.} Since $P \in \mathcal A$ then $P\mathcal A
P$ is a subalgebra of $\mathcal A$. Therefore correctability of
$\mathcal A$ implies correctability of $P\mathcal A P$ which from
Theorem~\ref{found} implies that $[P E_a^\dagger E_b^{} P, X] = 0$
for all $X \in P \mathcal A P$. Reciprocally, if this condition is
satisfied then by Theorem~\ref{found} there exists a channel
$\mathcal R$ correcting the algebra $P\mathcal A P$. In fact this
channel corrects all of $\mathcal A$. Indeed, remember that the
channel $\mathcal R$ built in the proof is such that $\mathcal
R^\dagger(X) = \mathcal R^\dagger(P X P)$ for all $X$. Therefore
for all $X \in \mathcal A$, $P(\mathcal R \circ
\channel)^\dagger(X)P = P (\mathcal R \circ
\channel)^\dagger(PXP)P = PXP$. \qed

\smallskip

In practice, the operation elements for a channel are usually not
known precisely; often it is just the linear space they span that
is known \cite{knill02}. Thus, for the explicit construction of
correction operations in Theorems~\ref{found} and \ref{found2} to
be of practical value, one has to show that the correction channel
$\mathcal R$ also corrects any channel whose elements are linear
combinations of the elements $E_a$. This is indeed the case. A
simple way to see this is to note that if the testable conditions
for conserved algebras of Theorems~\ref{cons} and \ref{cons2} are
satisfied for $\mathcal R \circ \mathcal E$, then they are also
satisfied for $\mathcal R \circ \mathcal E^\prime$ where the
operation elements of $\mathcal E^\prime$ are linear combinations
of those for $\mathcal E$.

\subsection{The Schr\"odinger Picture}\label{sS:schrodinger}

In order to illustrate how OAQEC generalizes OQEC we restate a special case of the above results 
in the Schr\"odinger picture: Suppose we have a decomposition
\begin{equation}
\Hil =  \Bigl[{\bigoplus_k \big( A_k \otimes B_k\big) }\Bigr]
\oplus \mathcal K,
\end{equation}
with $P$ the projector of $\Hil$ onto $\mathcal K^\perp =
\bigoplus_k A_k \otimes B_k$. The algebra in question includes $P$
as its unit and is given by $\mathcal A = \Big[ \oplus_k (\mathcal
L(A_k) \otimes \one_{B_k})\Big] \oplus 0_{\mathcal K}$. Observe
that the hypotheses of both results Theorems~\ref{found} and
\ref{found2} are satisfied. It follows that $\mathcal A$ is
correctable for $\mathcal E$ for states in $P\Hil$ if and only if
there exists a channel $\mathcal R$ such that for any density
operator $\rho = \sum_k \alpha_k (\rho_k \otimes \tau_k)$ with
$\rho_k \in \mathcal L_1(A_k)$, $\tau_k \in \mathcal L_1(B_k)$,
and nonnegative scalars $\sum_k \alpha_k=1$, there are operators
$\tau_k' \in \mathcal L_1(B_k)$ for which
\begin{equation}\label{Schroeqn}
(\mathcal R\circ\mathcal E)(\rho) = \sum_k \alpha_k \mathcal R
\bigl({\channel \bigl({\rho_k \otimes \tau_k}\bigr)}\bigr) =
\sum_k \alpha_k ( \rho_k \otimes \tau_k').
\end{equation}
Experimentally, each of the subsystems $A_k$ can be used
individually to encode quantum information. An extra feature of
this OAQEC code is the fact that an arbitrary mixture of encoded
states, one for each subsystem, can be simultaneously corrected by
the same correction operation.

By Theorem~\ref{found} (or Theorem~\ref{found2}), there is a
correction operation $\mathcal R$ for which Eq.~(\ref{Schroeqn})
is satisfied if and only if for all $a,b$ there are operators
$X_{abk} \in \mathcal L(B_k)$ such that
\begin{equation}
P E_a^\dagger E_b^{} P = \sum_k \one_{A_k} \otimes X_{abk}.
\end{equation}

Note that contrary to the Heisenberg formulation of
Eq.~(\ref{Heiseneqn}), the formulation of Eq.~(\ref{Schroeqn})
implicitly relies on the representation theory for
finite-dimensional C$^*$-algebras. As the representation theory
for arbitrary C$^*$-algebras is intractable, this suggests the
Heisenberg picture may be more appropriate for an
infinite-dimensional generalization of this framework.

Let us consider a qubit-based class of examples to illustrate the
equivalence established in Theorem~\ref{found}. A specific case
was discussed in \cite{beny07}. Suppose we have a hybrid quantum
code wherein $d$ qubit codes $\ket{\psi_j}$, $1\leq j \leq d$, are
each labelled by a classical ``address'' $\ket{j}$, $1\leq j \leq
d$. In this case $P = \sum_{j=1}^d \one_2\otimes \ket{j}\bra{j}  =
\one_2 \otimes \one_d$ and the algebra is $\mathcal A =
\oplus_{j=1}^d \mathcal L(\mathbf{C}^2) \otimes \ket{j}\bra{j}$. A
generic density operator for this code is of the form $\rho =
\sum_{j=1}^d \alpha_j \rho_j \otimes \ket{j}\bra{j}$, where
$\rho_j = \ket{\psi_j}\bra{\psi_j}$, $\alpha_j\geq 0$, and
$\sum_{j=1}^d \alpha_j =1$. This hybrid code determined by
$\mathcal A$ and $P$ is correctable for $\mathcal E$ if and only
if for all $a,b,j$ there are scalars $\lambda_{abj}$ such that
$$
PE_a^\dagger E_b P = \sum_{j=1}^d \lambda_{abj} (\one_2 \otimes
\ket{j}\bra{j}).
$$
As the ancilla for each individual qubit $\ket{\psi_j}$ is
one-dimensional, in this case the correction operation will
correct the code precisely, $(\mathcal R \circ \mathcal E)(\rho) =
\rho$.

In the Schr\"odinger picture, the correction channel built in the
proofs of Theorems~\ref{found} and \ref{found2}  is equal to
$$
\mathcal R(\rho) = \sum_k \mathcal R_k(Q_k\rho Q_k) +
\frac{\tr(Q^\perp \rho)}{\tr P} P,
$$
In words, one first measures the observable defined by the
complete set of orthogonal projectors $Q_k$ and $Q^\perp$. If the
state is found to be in one of the subspaces $Q_k$ then the OQEC
correction channel $\mathcal R_k$ is applied to correct the
corresponding subsystem of the algebra. Otherwise, if the state
happens to be in the subspace $Q^\perp$, this means that the
initial state of the system was not in the code. Therefore what we
do in this case does not matter. In the channel $\mathcal R$
defined above, we chose for simplicity to set the state to $P/\tr
P$.


\section{Application to Information Flow}\label{S:informationflow}

Consider the interaction between a ``system'' $S$ and an
``apparatus'' $A$ where the initial state of the apparatus is
known. For any state $\ket{\psi_S} \in \Hil_S$, we define
$V\ket{\psi_S} \doteq U(\ket{\psi_S}\otimes \ket{\psi_A}) $ for a
unitary $U$ acting on $\Hil_S \otimes \Hil_A$ and a fixed initial
state $\ket{\psi_A} \in \Hil_A$. The operator $V$ is an isometry
between the space $\Hil_S$ and the space $\Hil_S \otimes \Hil_A$.
Tracing over the final state of the apparatus gives us a channel
from $B(\Hil_S)$ to $B(\Hil_S)$: $ \channel_{SS}(\rho) =
\tr_A(V\rho V^\dagger)$ whose dual is
$$
\channel_{SS}^\dagger(X) = V^\dagger(X \otimes \one) V
$$
\begin{figure}
\begin{center}
\includegraphics[width=0.3\columnwidth]{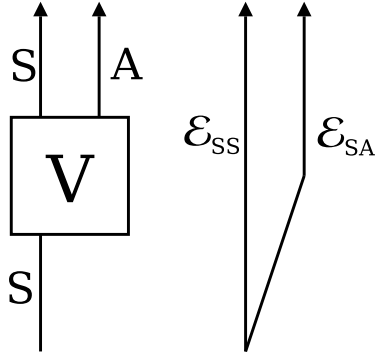}
\caption{\small Interaction between a system $S$ and an apparatus
$A$ of known initial state. Tracing over one of the two final
system gives us one of two channels $\channel_{SS}$ or
$\channel_{SA}$.} \label{fig:channels}
\end{center}
\end{figure}
We can also trace out the final state of the system to get a
channel from $\mathcal L(\Hil_S)$ to $\mathcal L(\Hil_A)$: $
\channel_{SA}(\rho) = \tr_S(V\rho V^\dagger) $ where $\rho \in
B(\Hil_S)$. See Figure~\ref{fig:channels}. The channel
$\channel_{SA}$ is uniquely defined by $\channel_{SS}$, up to an
arbitrary unitary operation on the apparatus, and is usually
called the {\em complementary channel} of $\channel_{SS}$. Its
dual is simply
$$
\channel_{SA}^\dagger(Y) = V^\dagger(\one \otimes Y) V.
$$

Using Theorem~\ref{found}, we can determine which observables have
been preserved by either $\channel_{SS}$ or $\channel_{SA}$,
irrespectively of the system's initial state.
The answers are given by two subalgebras of $\mathcal L(\Hil_S)$:
respectively $\mathcal A_{SS}$ and $\mathcal A_{SA}$. The algebra
$\mathcal A_{SS}$ characterizes the information about the system's
initial state which has been preserved by the system's evolution,
and $\mathcal A_{SA}$ characterizes the information about the
system's initial state which has been transferred to the
environment.

Those  algebras can be expressed in terms of the elements of one
of the channels. For instance, if $E_a$ are elements for
$\channel_{SS}$, then $V$ can be expressed as $ V = \sum_b E_b
\otimes \ket{\phi_b^A} $ for some orthonormal set of vectors
$\ket{\phi_b^A}$ of $\Hil_A$. Hence for any choice of a basis
$\ket{a}$ of $\Hil_S$ we obtain a family of elements for the
channel $\channel_{SA}$, namely
$$
F_a = \sum_b \ket{\phi_b^A} \bra{a} E_b.
$$
This means that the relevant operators entering
Theorem~\ref{found} for the second channel are
$$ F_a^\dagger F_b^{} =
\sum_c E_c^\dagger\ketbra{a}{b}E_c^{} =
\channel_{SS}^\dagger(\ketbra{a}{b}).
$$
Note that the operators $\ketbra{a}{b}$ form a basis for the whole
operator algebra $\mathcal L(\Hil_S)$. Hence the observables
correctable for the apparatus form the algebra $ \mathcal A_{SA} =
{\rm Alg}\bigl( {\rm Ran}\, \channel_{SS}^\dagger \bigr)' $: the
algebra of operators commuting with all operators in the range of
$\channel_{SS}^\dagger$.
Hence we see that a direct consequence of Theorem~\ref{found} is
that in an open dynamics defined by a channel $\channel$, full
information about a projective observable can escape the system if
and only if it commutes with the range of the dual map
$\channelb$, which is the set of observables with first moment
information conserved by $\channel$. This generalizes results in
\cite{lindblad99}.

We can characterize the observables representing information which
has been ``duplicated'' between the system and the apparatus. They
form the intersection
$$
\mathcal C := \mathcal A_{SS} \cap \mathcal A_{SA}.
$$
From the correctability of $\mathcal A_{SS}$
(Eq.~(\ref{Heiseneqn})) we have that $\mathcal A_{SS} \subseteq
{\rm Ran}\, \channel_{SS}^\dagger$, from which it follows that
${\mathcal A}_{SA} = {\rm Alg}\bigl( {\rm Ran}\,
\channel_{SS}^\dagger \bigr)' \subseteq \mathcal A_{SS}'$. Hence,
the algebra of duplicated observables is $\mathcal C \subseteq
\mathcal A_{SS}' \cap \mathcal A_{SS}$, where $\mathcal A_{SS}'
\cap \mathcal A_{SS}$ is the center of $A_{SS}$: those elements of
the algebra which commute with all other elements. In particular,
the duplicated algebra $\mathcal C$ is {\em commutative}. Note
that the contrary would have violated the no-cloning theorem after
correction of both channels. Since the algebra $\mathcal C$ is
commutative, it is generated by a single projective observable
which can be represented by a self-adjoint operator $C$.

Given that, after the interaction, both the system and the
apparatus contain information about the same observable $C$ on the
initial state of the system, we may expect that they are
correlated. Let $P_i \in \mathcal C$ be the projectors on the
eigenspaces of $C$. There exists a POVM with elements $X_i$ on the
system as well as a POVM $Y_i$ on the apparatus such that
$$
\channelb_{SS}(X_i) = \channelb_{SA}(Y_i) = P_i.
$$
Note that if $\mathcal R_{SS}$ and $\mathcal R_{SA}$ are
correction channels for $\channel_{SS}$ and respectively
$\channel_{SA}$, then $X_i = \mathcal R_{SS}^\dagger(P_i)$ and
$Y_i = \mathcal R_{SA}^\dagger(P_i)$.

We will show that the observables $X_i$ and $Y_i$ are correlated.
First note that $\tr(P_i P_k) = \delta_{ki}$, which we can also
write
$$
\tr(P_i P_k) = \tr(P_i \channelb_{SS}(X_k)) = \tr(P_i V^\dagger
(X_k \otimes \one) V) = \delta_{ki}
$$
 This means that when $k \neq i$, $(X_k \otimes \one) V P_i = 0$, which can be seen by expending $P_i$ in terms of eigenvectors. Also since $\sum_k X_k = \one$, then $V P_i = (X_i \otimes \one) V P_i$. The same argument is true also for $Y_k$. Therefore $(X_k \otimes \one) V P_i = (\one \otimes Y_k) V P_i = \delta_{ik} V P_i$. Hence
\[
\begin{split}
V^\dagger (X_i \otimes Y_j) V &= \sum_{kl} P_k V^\dagger (X_i
\otimes Y_j) V P_l = \sum_{kl} \delta_{ik} \delta_{jl} P_i P_j =
\delta_{ij} P_k
\end{split}
\]
which means that for any state $\rho$ of the system
$$
\tr(V\rho V^\dagger (X_i\otimes Y_j)) = \delta_{ij} \tr(\rho P_i).
$$
Hence the probability that the outcome of a measurement of $X$
differs from that of $Y$ is zero. This means that the information
that the apparatus ``learns'' about the system and which is
characterized by the observable $C$ is correlated with a property
of the system after the interaction. Therefore $C$ represents the
only information that the apparatus acquires about the system and
which stays pertinent through the interaction.

This analysis has implication for the theory of decoherence
\cite{giulini96, zurek03} as well as for the theory of
measurements. We have shown that any interaction between a system
and its environment (which took the role of the apparatus)
automatically selects a unique observable $C$ as being the only
{\em predictive} information about the system acquired by the
environment. Even though an observer who has access to the
environment could learn about any observable contained in the
algebra $\mathcal A_{SA}$, only the information encoded by $C$
bears any information about the future state of the system. This
suggests that the pointer states which characterize decoherence
should not be selected only for their stability under the
interaction with the environment: One should also add the
requirement that they encode information that the environment
learns about the system. Indeed any one of those requirements
taken separately does not select a single observable
unambiguously, but together they do. This is a new way of solving
the {\em basis ambiguity} problem \cite{zurek81}.



\section{Error Correction of Operator Spaces}\label{S:operatorspaces}

In this section we discuss an extension of OAQEC theory to the
setting of operator spaces generated by observables. We shall
leave a deeper analysis of this extension for investigation
elsewhere. An {\it operator space} \cite{Pau02} is a linear
manifold (a subspace) of operators inside $\mathcal L(\mathcal
H)$. Operator spaces, and their Hermitian closed counterpart
``operator systems'', have arisen recently in the study of channel
capacity problems in quantum information \cite{devetak05}. Observe
that (by design) Definitions~\ref{conserved} and \ref{correctable}
include the case of operator spaces and systems generated by
observables, and hence these cases fit into the mathematical
framework for error correction introduced here. Let us describe
how operator spaces physically arise in the present setting.

In Section \ref{sS:testable} we showed how to build the correction
channel for active error correction. We were free to choose what
to do to the system in the case that the syndrome measurement
revealed the state had not initially been in the code prior to the
action of the error channel. In fact, there is an advantage in
choosing to send that state back in the code, meaning that we
choose the correction channel such that $\mathcal R^\dagger(X) =
\mathcal R^\dagger(PXP)$ for any operator $X$, which is indeed the
case for the correction channel defined in Theorem~\ref{found2}.
Consider the set of operators defined by
$$
\mathcal V := \channelb( \mathcal R^\dagger (\mathcal A)).
$$
This set is not an algebra in general. Nevertheless, it is an
operator system by the linearity and positivity of channels. If $X
\in \mathcal V$ then there exists $Y \in P\mathcal AP$ such that
$X = \channelb( \mathcal R^\dagger (Y))$ and also $PXP = Y$.
Therefore for all $X \in \mathcal V$,
$$
\channelb( \mathcal R^\dagger (X)) = \channelb( \mathcal R^\dagger
(Y)) = X.
$$
Hence the observables in $\mathcal V$ are exactly corrected, and
this is {\it independent} of what the initial state was. For
instance, if we ``forgot'' to make sure that the initial state was
in the code, we can still recover all the information, provided
that we measure the observable with elements $X_k = \channelb(
\mathcal R^\dagger (Y_k))$ whenever we would have measured $Y_k
\in \mathcal A$. Typically this would involve measuring general
(unsharp) POVMs instead of just sharp projective observables. This
shows that it could be useful to consider the correction of
general POVMs. Since POVM elements do not always span an algebra,
this suggests that we should consider the correctability (passive
or active) codes associated with operator systems in this way.

\strut

Consider the following simple example of a conserved operator
space that is not an algebra. Let $\mathcal H$ be single qutrit
Hilbert space with computational basis
$\{\ket{0},\ket{1},\ket{2}\}$. Consider the channel $\mathcal E$
on $\mathcal H$ defined by its action on observables represented
in this basis as follows:
\begin{equation}
\mathcal E^\dagger \Big( [a_{ij}]_{3\times 3}  \Big) \quad = \quad
\left[
\begin{matrix} a_{11} & a_{12} & 0 \\ a_{21} & a_{22} & 0 \\ 0 & 0
& \frac{a_{11}+a_{22}}{2}
\end{matrix} \right].
\end{equation}
Observe that $(\mathcal E^\dagger)^2 = \mathcal E^\dagger \circ
\mathcal E^\dagger = \mathcal E^\dagger$ and that the range
$\mathcal V$ of $\mathcal E^\dagger$ coincides with its fixed
point set; $\{ Y :  Y = \mathcal E^\dagger (X)\,\,{\rm
for\,\,some\,\,}X \} = \{X:\mathcal E^\dagger (X)=X\}$. Thus, the
operator system $\mathcal V$ is conserved by $\mathcal E$.
Moreover, $\mathcal V$ is not an algebra since it is not closed
under multiplication.

\strut

Given results from other settings for quantum error correction, it
is of course desirable to find a characterization of correction
for operator spaces independent of any particular recovery
operation. Here we derive a necessary condition, and we leave the
general question as an open problem. Observe that if there exists
a channel $\mathcal R$ such that $P\channelb( \mathcal R^\dagger
(X))P = PXP$ for all $X \in \mathcal V$ then, $0 \le X \le \one$
implies $0 \le \mathcal R^\dagger (X) \le \one$, since $\mathcal
R^\dagger$ is a contractive map. This in turn implies that there
exists $0 \le Y \le \one$ such that $PXP = P\channelb(Y)P$;
namely, $Y=\mathcal R^\dagger(X)$.

\begin{prop} \label{prop:opspacecond}
A necessary condition for an operator space $V$ to be correctable
on states $P\Hil$ for $\channel$ is that for all $X \in \mathcal
V$ such that $0 \le X \le \one$, there exists $0 \le Y \le \one$
such that $PXP = P\channelb(Y)P$. This condition is also
sufficient when $\mathcal V$ is an algebra containing $P$.
\end{prop}

\Prf The first part of the statement has been proved. We show that
the condition expressed implies correctability of $\mathcal V$ if
it is an algebra containing $P$.

Since $P \in \mathcal V$, we know that $\mathcal B := P\mathcal V
P$ is a subalgebra of $\mathcal V$. Note it is sufficient to prove
the correctability of $\mathcal B$. Indeed, if $\mathcal B$ is
correctable then in particular $P$ is correctable. Let $\mathcal
R$ be a correction channel for the largest correctable algebra on
$P\Hil$. Then we have seen in the proof of Theorem~\ref{found2}
that the correctability of $P$ implies that $P\channelb(\mathcal
R^\dagger(X))P = P\channelb(\mathcal R^\dagger(P X P))P$ for any
$X$. From this it follows that the correctability of $\mathcal B$
on $P\Hil$ implies that of $\mathcal V$.

Let ${\mathbb E} := \{X \;|\; 0 \le X \le \one\}$. Suppose that $P
\mathcal (V \cap {\mathbb E}) P = \mathcal B \cap {\mathbb E}
\subseteq P\channelb({\mathbb E})P$. Then for all projectors $Q
\in \mathcal B$, there exists a self-adjoint operator $0 \le X \le
\one$ such that $P\channelb(X)P = Q$. Hence
$(P-Q)\channelb(X)(P-Q) = 0$, which implies $X E_k (P-Q) = 0$, or
$X E_k P = X E_k Q$ for all $k$. Also we have $P\channelb(\one -
X)P = P - Q$, so that $Q\channelb(\one - X)Q = 0$ from which
$(\one - X)E_k Q = 0$, or $E_k Q = X E_k Q$ for all $k$. Therefore
$$
X (E_k P) = X E_kQ = E_k Q = (E_k P) Q.
$$
Combining the two results we obtain
$$
Q P E_k^\dagger E_j P = P E_k^\dagger X E_j P = P E_k^\dagger E_j
P Q
$$
for all $k$ and all $j$. This result can linearly be extended to
the whole of $\mathcal B$, since an algebra is spanned by its
projectors. Hence $[X, P E_k^\dagger E_j P] = 0$ for all $X \in
\mathcal B$. From Theorem~\ref{found2} this implies that $\mathcal
B$, and therefore the algebra $\mathcal V$ is correctable on
$P\Hil$. \qed

\strut

As a final example, consider the noise model corresponding to a
single random bit flip on three qubits. The noise operators are
$\{\one, X_1, X_2, X_3\}$ where $X_i$ is a Pauli $x$ matrix on the
$i$th qubit. One can correct the standard quantum code with
projector $P = \proj{000} + \proj{111}$ expressed in the
computational basis. It is easy to check that a correction channel
for this code is
$$
\mathcal R^\dagger(A) = PAP + \sum_i X_i PAP X_i
$$
This channel has the properties that we need in order to ``lift''
the code to an operator space code correctable on all states.
Indeed, we have $\mathcal R^\dagger(P) = \one$ and $\mathcal
R^\dagger(A) = \mathcal R^\dagger(PAP)$ for all $A$. The algebra
correctable on the code $P$ is
$$
\mathcal A = \Bigl\{ \sum_{ij} \alpha_{ij} \ket{iii}\bra{jjj} :
\alpha_{ij} \in \mathbb C \Bigr\}
$$
In order to proceed however we need a specific error channel,
which is given by choosing a probability for the occurrence of
each error:
$$
\channel(\rho) = p_0 \rho + \sum_i p_i X_i \rho X_i
$$
Then, writing $X_0 = \one$, the operator space
$$
\mathcal V = (\mathcal R \circ \channel)^\dagger(\mathcal A) =
\Bigl\{ \sum_{i,j=0}^1 \alpha_{ij} \sum_{k,l = 0}^3 p_{k} X_k X_l
\ket{iii}\bra{jjj} X_l X_k : \alpha_{ij} \in \mathbb C \Bigr\}
$$
is correctable by $\mathcal R$ on all states. We should have $P
\mathcal V P = \mathcal A$. This can be seen from the fact that
$\ket{iii} = P \ket{iii}$ and $P X_k X_l P = \delta_{kl} P$.
Explicitly separating the components respectively inside $\mathcal
A$ and orthogonal to $\mathcal A$ we have
operators in $\mathcal V$ which live in $\mathcal A$ and outside
of
$\mathcal A$:
$$
\mathcal V = \Bigl\{ \sum_{ij} \alpha_{ij} \Bigl({
\ket{iii}\bra{jjj} + \sum_{k \neq l} p_{k} X_k X_l
\ket{iii}\bra{jjj} X_l X_k }\Bigr) : \alpha_{ij} \in \mathbb C
\Bigr\}
$$




\vspace{0.1in}

{\noindent}{\it Acknowledgements.} We are grateful to the Banff
International Research Station for kind hospitality. Some of the
work for this paper took place during workshop 07w5119 on
``Operator structures in quantum information theory''. This work
was partially supported by NSERC, PREA, ERA, CFI, OIT, and the
Canada Research Chairs program.


\end{document}